\documentclass{iau379}

\usepackage{natbib}
\usepackage{graphicx}
\usepackage{amsmath}
\usepackage{multirow}
\usepackage{journals}

\begin{document}

\lefttitle{Chilingarian et al.}
\righttitle{Dynamical masses of UDGs}

\jnlPage{1}{7}
\jnlDoiYr{2023}
\doival{10.1017/xxxxx}

\aopheadtitle{Proceedings of IAU Symposium 379}
\editors{P. Bonifacio,  M.-R. Cioni, F. Hammer, M. Pawlowski, and S. Taibi, eds.}

\title{Dark matter content and dynamical masses of ultra-diffuse galaxies in the Coma cluster}

\author{Igor V. Chilingarian $^{1,2}$, Kirill A. Grishin$^{3,2}$, Anton V. Afanasiev$^{4,2}$,
Anton Mironov$^{5}$, Daniel Fabricant$^{1}$, Sean Moran$^{1}$, Nelson Caldwell$^{1}$, Ivan Yu. Katkov$^{6,2}$, Irina Ershova$^{5}$}

\affiliation{$^1$Center for Astrophysics -- Harvard and Smithsonian, 60 Garden St. Cambridge, MA, 02138 USA\\
$^2$Sternberg Astronomical Institute, Moscow State University, 13 Universitetsky pr., Moscow, Russia\\
$^3$Universit\'e Paris Cit\'e, CNRS, Astroparticule et Cosmologie, F-75013 Paris, France\\
$^4$LESIA, Observatoire de Paris, 5 place Jules Janssen, 92195, Meudon, France\\
$^5$Faculty of Space Research, Moscow State University, 1 Leninskie Gory, bld.~52, Moscow, Russia\\
$^6$New York University Abu Dhabi, Saadiyat Island, PO Box 129188, Abu Dhabi, UAE}

\begin{abstract}
Ultra-diffuse galaxies (UDGs) are spatially extended, low surface brightness stellar systems with regular elliptical-like morphology found in large numbers in galaxy clusters and groups. Studies of the internal dynamics and dark matter content of UDGs have been hampered by their low surface brightnesses.  We identified a sample of low-mass early-type post-starburst galaxies, `future UDGs' in the Coma cluster still populated with young stars, which will passively evolve into UDGs in the next 5--10 Gyr.  We collected deep observations for a sample of low-mass early-type galaxies in the Coma cluster using MMT Binospec, which includes present-day and future UDGs. We derived their dark matter content within a half-light radius (70--95~\%) and total dynamical masses ($M_{200}=5.5\cdot10^9-1.4\cdot10^{11} M_{\odot}$) assuming the Burkert density profile and assess how different proposed evolutionary channels affect dark and visible matter in UDGs. We also discuss observational methodology of present and future UDG studies.
\end{abstract}


\maketitle

\section{Introduction}

Ultra-diffuse galaxies (UDGs) are low-surface brightness ($\langle \mu_r \rangle>24$~mag~arcsec$^{-2}$) extended ($R_e$ up-to 10~kpc) stellar systems.  Initially discovered in the Virgo cluster in the 80s \citep{SB84}, they were nearly forgotten for almost two decades until \citet{2006ApJ...651..822C} resolved red giant branch stars in two of them.  Another decade later, UDGs attracted a lot of attention, when hudreds of them were discovered in the Coma cluster \citep{2015ApJ...798L..45V,Koda15} and progress in astronomical instrumentation enabled their spectroscopic studies in integrated light.  The main unanswered question is whether UDGs represent an extension of dwarf elliptical (dE) galaxy population to larger sizes and lower surface brightnesses \citep{2018RNAAS...2a..43C} and share common evolutionary paths, or whether they form a separate galaxy class with different formation scenarios \citep[see e.g.][]{2015MNRAS.452..937Y}.

The proposed UDG formation scenarios can be confirmed or ruled out by detailed assessment of their internal dynamics, dark matter content, and  star formation histories. Here we review our recent results on dynamical modelling of UDGs and put them into the context of their formation and evolution.

\section{UDG Observations and Data Reduction: Methodological Considerations}

Dynamical modelling of collisionless stellar systems requires knowledge of the stellar density distribution and internal kinematics (stellar velocities, velocity dispersions, possibly higher order moments of the line-of-sight velocity distribution (LOSVD), and proper motions of individual stars for nearby galaxies and star clusters).  Therefore, an input observational dataset usually contains images and spectra, or, in rare cases, a fully  calibrated spectral data cube, which can be used for stellar density measurements, even though direct images are typically more reliable in the areas of low surface brightness.

\subsection{Imaging and Light Profile Decomposition}

\begin{figure}
    \centering
    \includegraphics[width=0.48\hsize]{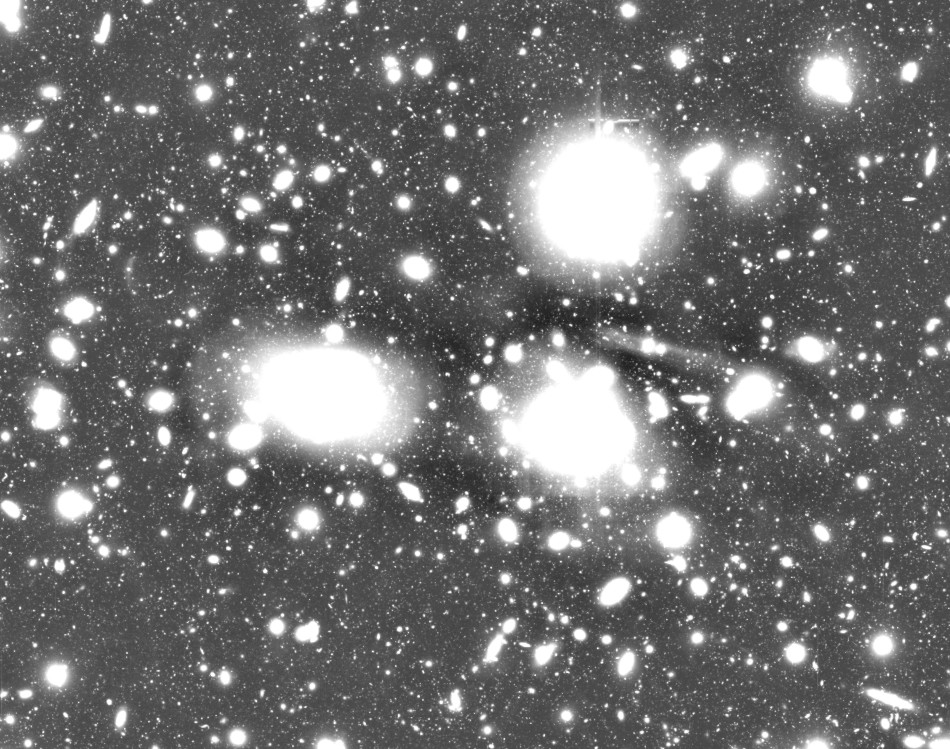}
    \includegraphics[width=0.48\hsize]{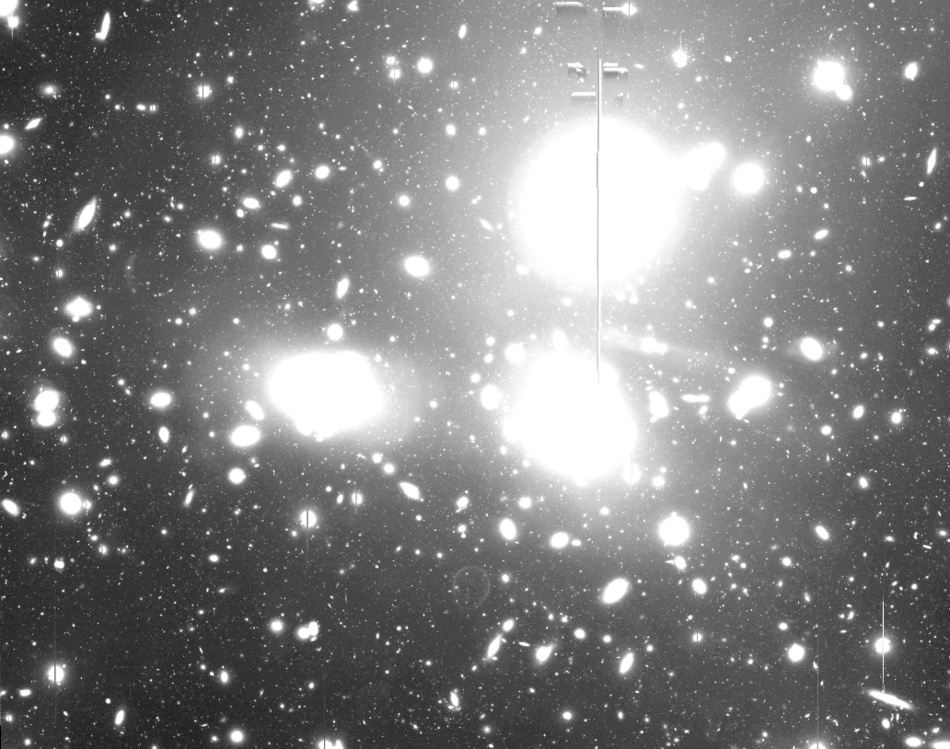}
    \caption{Improvement in sky background subtraction for HyperSuprimeCam data for the central region of the Coma cluster. We show contrast-enhanced images obtained from the HyperSuprimeCam Legacy Archive (left) and using our own data reduction (right). The background over-subtraction in the left panel reults in dark rings around the two dominant cluster galaxies.
    } \label{fig:coma_image}
\end{figure}

When observed from the ground, UDG central surface brightnesses fall below that of the dark moonless sky, and one typically needs to trace light profiles to $\mu_r>26$~mag~arcsec$^{-2}$ to confidently measure an effective radius and S\'ersic concentration.  Therefore, the quality of sky subtraction is the key quality factor for light profiles.  Unfortunately, standard mosaic imaging data processing pipelines use local sky background modelling, leading to background oversubtraction near extended targets and to artefacts in galaxies close to the edges of individual detectors in the array.   For this project, we modified the Subaru HyperSuprimeCam data reduction pipeline \citep{2018PASJ...70S...5B} to construct a global sky background model (see Fig.~\ref{fig:coma_image}), which allowed us to mitigate these problems and confidently reach the surface brightnesses $\mu_g=28.5\dots29.0$~mag~arcsec$^{-2}$.

Relatively nearby galaxies (to a distance of $\sim$20~Mpc) can be resolved into stars using Hubble and the James Webb Space telescopes. Star counts allow a more precise determination of the outer light profiles. Because resolved images don't reach the faintest stars we need to renormalize the star count density profile using ground-based data \citep{2023MNRAS.520.6312A}.  Even though crowding might become an issue in the inner parts of a galaxy we are able to define a region where both unresolved and resolved profiles overlap for renomalization.  At the Virgo cluster distance ($\sim$16~Mpc) at a surface brightness of $\mu_r=23.5 \dots 26.0$~mag~arcsec$^{-2}$ confusion does not affect the population of red giant stars.

\subsection{Optical Spectroscopy}
Obtaining integrated-light spectra of UDGs and using them to extract stellar kinematics, ages, and metallcities remains challenging.  However, even relatively low signal-to-noise spectra (S/N$\sim$3~pix$^{-1}$) can yield good measurements of internal dynamics with an appropriate observing strategy \citep{Chilingarian+19}.

The key factors are spectral resolution, choice of wavelength range, and slit width for long-slit or multi-slit observations.  Three main rules apply. (1) The slit width should not exceed the seeing FWHM; otherwise the galaxy light profile will define the spectral line shape making internal kinematics measurements impossible. (2) The spectral resolution ($\sigma_{\mathrm{inst}}$) should match the expected velocity dispersion of the galaxy ($R\approx5000$ for UDGs). If the resolution is too low measuring velocity dispersions and radial velocities will be difficult or impossible. If the resolution is too high, the LOSVD will be oversampled, leading to no improvement in $v_*$ and $\sigma_*$ uncertainties. (3) The chosen wavelength range must contain many strong absorption lines. This selection affects the quality of kinematics more than any other factor \citep{2020PASP..132f4503C}.  Additionally, to make stellar population measurements possible, one should include age- (e.g.  Balmer lines) and metallicity-sensitive features in the spectral range, even though many more spectral lines can be used to determine stellar ages \citep{2009MNRAS.394.1229C} than typically understood. The best (optical) wavelength range for studies of internal kinematics and stellar populations of faint quiescent galaxies is between 3900--5400~\AA.

\section{Observations, Data Reduction and Analysis}

We observed UDGs and diffuse post-starburst galaxies (PSGs) or `future UDGs' in the Coma$=$Abell~1656 ($z=0.023$) and Abell~2147 ($z=0.035$) galaxy clusters using the high-throughput multi-object optical spectrograph Binospec \citep{2019PASP..131g5004F} at the f/5 focus of the 6.5~m MMT. 
The wavelength range $3700<\lambda<5300$~\AA\ at a spectral resolving power $R=4800$  was chosen to target low-surface brightness absorption-line galaxies. We used deep archival imaging of the Coma cluster from HyperSuprimeCam at the 8-m Subaru telescope.

We reduced spectroscopic data using the standard Binospec data reduction pipeline \citep{2019PASP..131g5005K}.  We reduced HSC images using a customized version of the HSC data reduction pipeline were we added a block to construct a global sky model.

\begin{figure}
    \centering
    \includegraphics[width=\hsize]{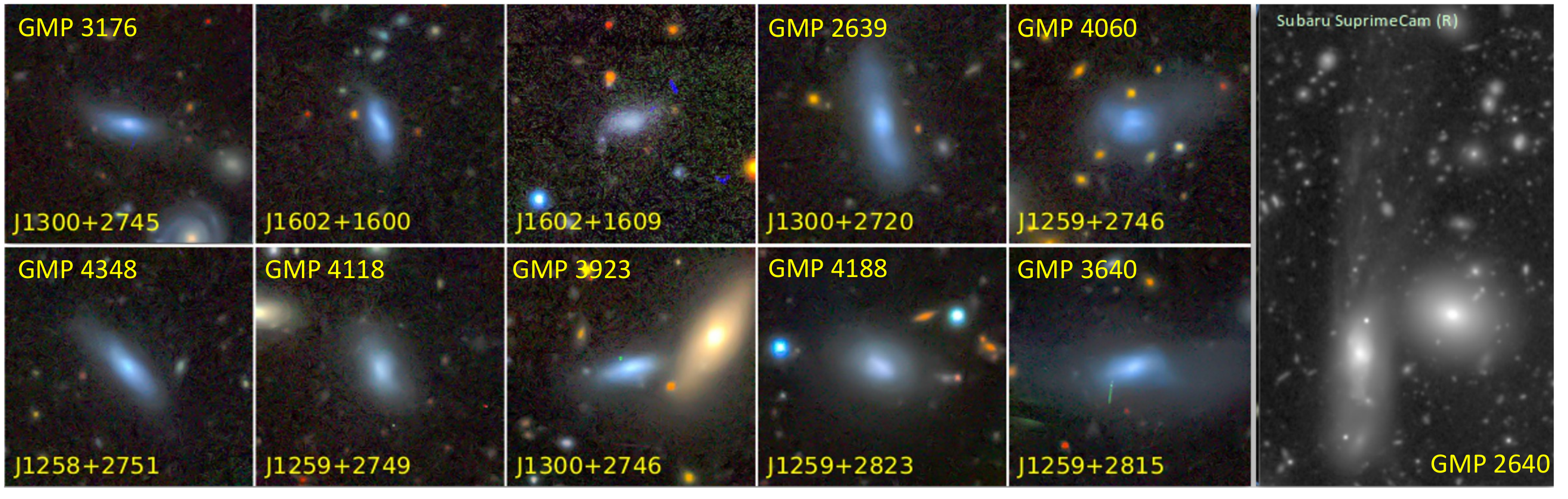}
    \caption{A sample of diffuse PSGs (8 `future UDGs' and 2 `future dEs' in the Coma and Abell~2147 galaxy clusters, for which we performed full dynamical modelling using internal kinematics out to $1.5-3 R_e$. Color images are assembled from archival Canada-France-Hawaiian Telescope images. For GMP~2640 we show a Subaru Suprimecam $R$-band image.}
    \label{fig:psg_sample}
\end{figure}

We use the {\sc NBursts} full spectum fitting code \citep{CPSK07} with $R=10000$ simple stellar population models computed with the {\sc pegase.hr} code \citep{LeBorgne+04} and a dedicated grid of synthetic spectra to model a ram-pressure-induced starburst \citep{2019arXiv190913460G} for diffuse PSGs (Fig.~\ref{fig:psg_sample}).  We analyze images with {\sc galfit} \citep{2010AJ....139.2097P} using up to three light profile components.  We perform Jeans axisymmetric modelling including a spherical dark matter halo in addition to the stellar component \citep{2023MNRAS.520.6312A}. 

\section{Dynamical Masses of UDGs}

Full details of the Binospec data analysis for the `bona fide UDGs' and for diffuse PSGs are presented in \citet{Chilingarian+19} and \citet{2021NatAs...5.1308G} respectively.  Here we discuss the derived dark matter content within $R_e$, how it is related to full (extrapolated) dynamical masses, and how assumptions about the dark matter profiles affect this extrapolation.

The derived dark matter content within 1~$R_e$ (70--95~\%) derived for UDGs using sparse measurements with $\sigma_*$ profiles including between 1 and 5 data points and extending barely to 1~$R_e$, agree remarkably well with the measurements of `future UDGs' from well sampled kinematics extending to 1.5-3~$R_e$. There seems to be a trend of increased DM for smaller stellar surface densities but this trend should be verified with a larger sample. The agreement between the two approaches validates the low-S/N measurements and proves that diffuse PSGs are indeed `future UDGs' despite their current high surface brightness arising from abundant young stars.

At the same time, the total virial masses $M_{200}$ derived under assumptions of NFW \citep{1997ApJ...490..493N} and \citet{Burkert95} DM profiles derived from stellar kinematics within the central 1--3~$R_e$ are different by up-to an order of magnitude (Burkert profiles being lighter). This phenomenon is best illustrated by the case of KDG~64, a small UDG or a large dwarf spheroidal galaxy in the nearby M~81 group where despite very low surface brightness, the profiles of internal kinematics are well sampled within 1~$R_e$ \citep{2023MNRAS.520.6312A}.

\begin{figure}
    \centering
    \includegraphics[width=0.48\hsize]{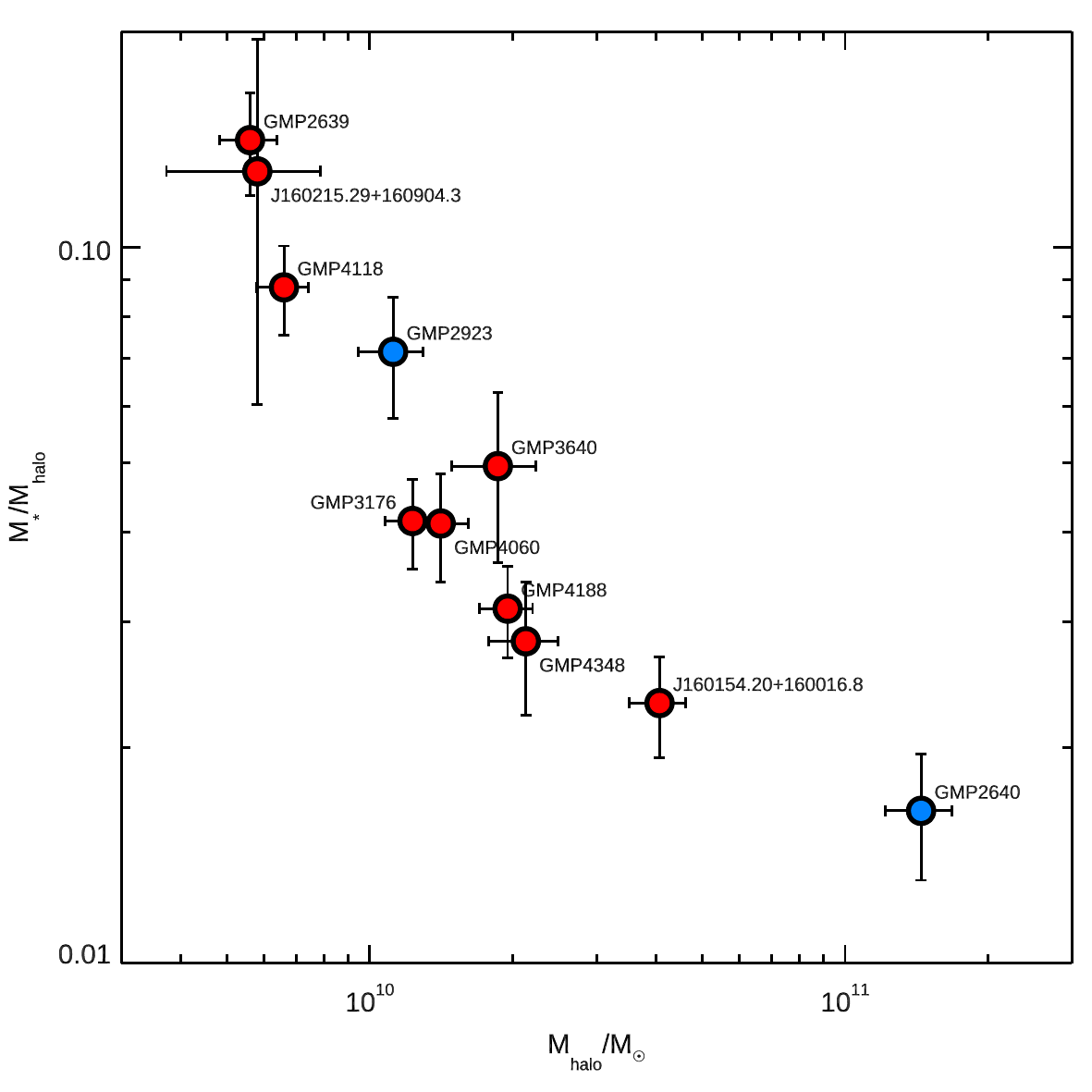}
    \includegraphics[width=0.48\hsize]{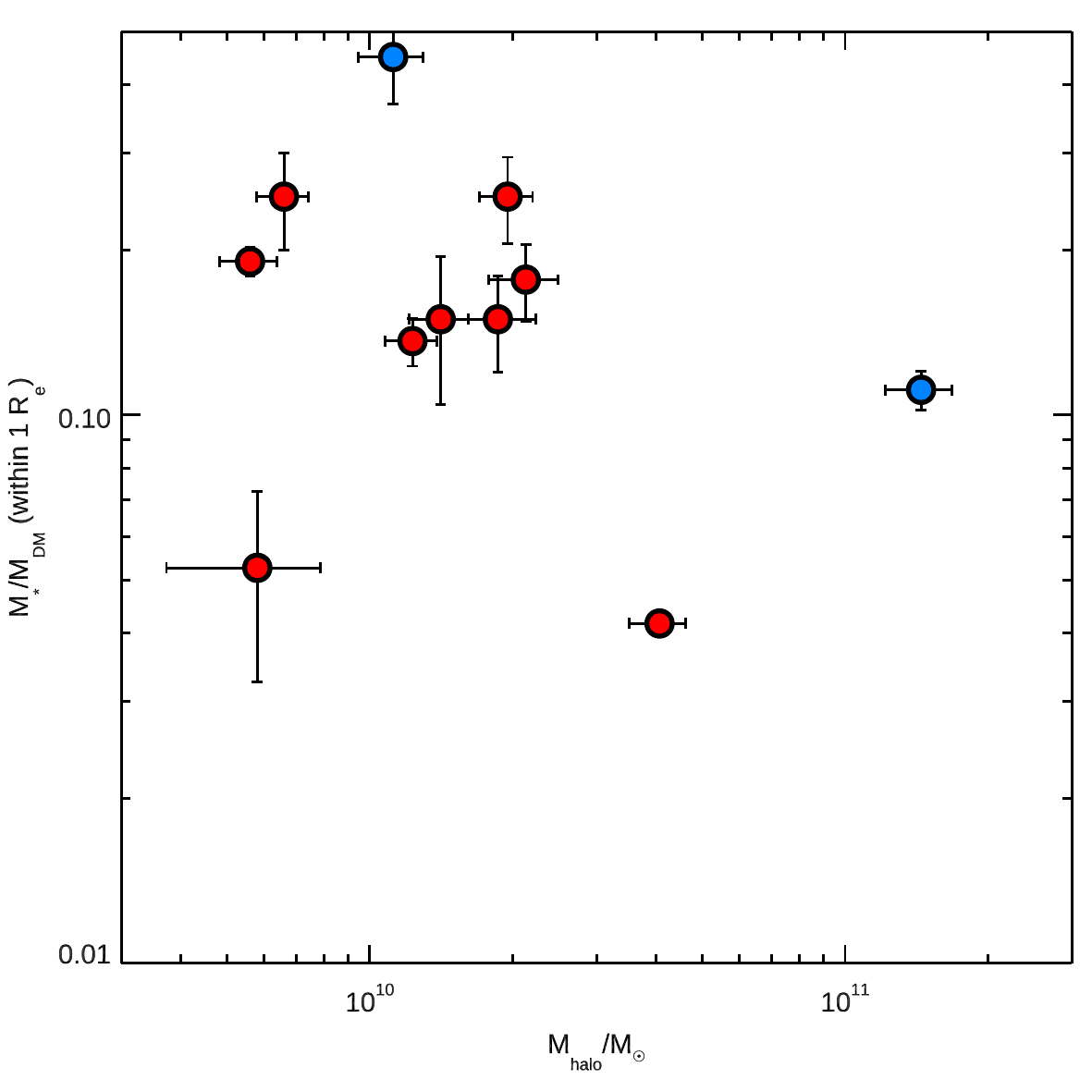}
    \caption{\textit{Left:} Total stellar-to-halo mass ratio vs total extrapolated halo mass for a sample of diffuse PSGs in Coma and Abell~2147. \textit{Right:} Stellar-to-dark matter mass ratio within $R_e$ vs total extrapolated halo mass. Red and blue dots mark galaxies, which will passively evolve into UDGs and dEs respectively.}
    \label{fig:sfe_udg}
\end{figure}

Using the measurements presented in \citet{2021NatAs...5.1308G} we can compare total stellar mass $M_*$ to a total dark matter halo mass $M_{200}$ computed for the Burkert profile, and perform the same comparison for the quantities within $R_e$. Interestingly, there seems to be a clear trend of the total $M_*/M_{200}$ ratio as a function of $M_{200}$ while no correlation is seen for the $M_*/M_{DM}$ within $R_e$ (see Fig.~\ref{fig:sfe_udg}). The total mass ratio trend goes in the opposite direction from expectation for the low-mass end of the galaxy mass function in the `star formation efficiency' diagram from \citet{2019MNRAS.488.3143B}.  For our sample of `future UDGs' the stellar-to-halo mass ratio increases when the halo mass decreases. We also notice that our values of $M_*/M_{\mathrm{halo}}$ are over an order of magnitude higher than those from \citet{2019MNRAS.488.3143B}.  This discrepancy may be reduced if we use a NFW halo density profile.

\section{Discussion}

\begin{figure}
    \centering
    \includegraphics[width=1.0\hsize]{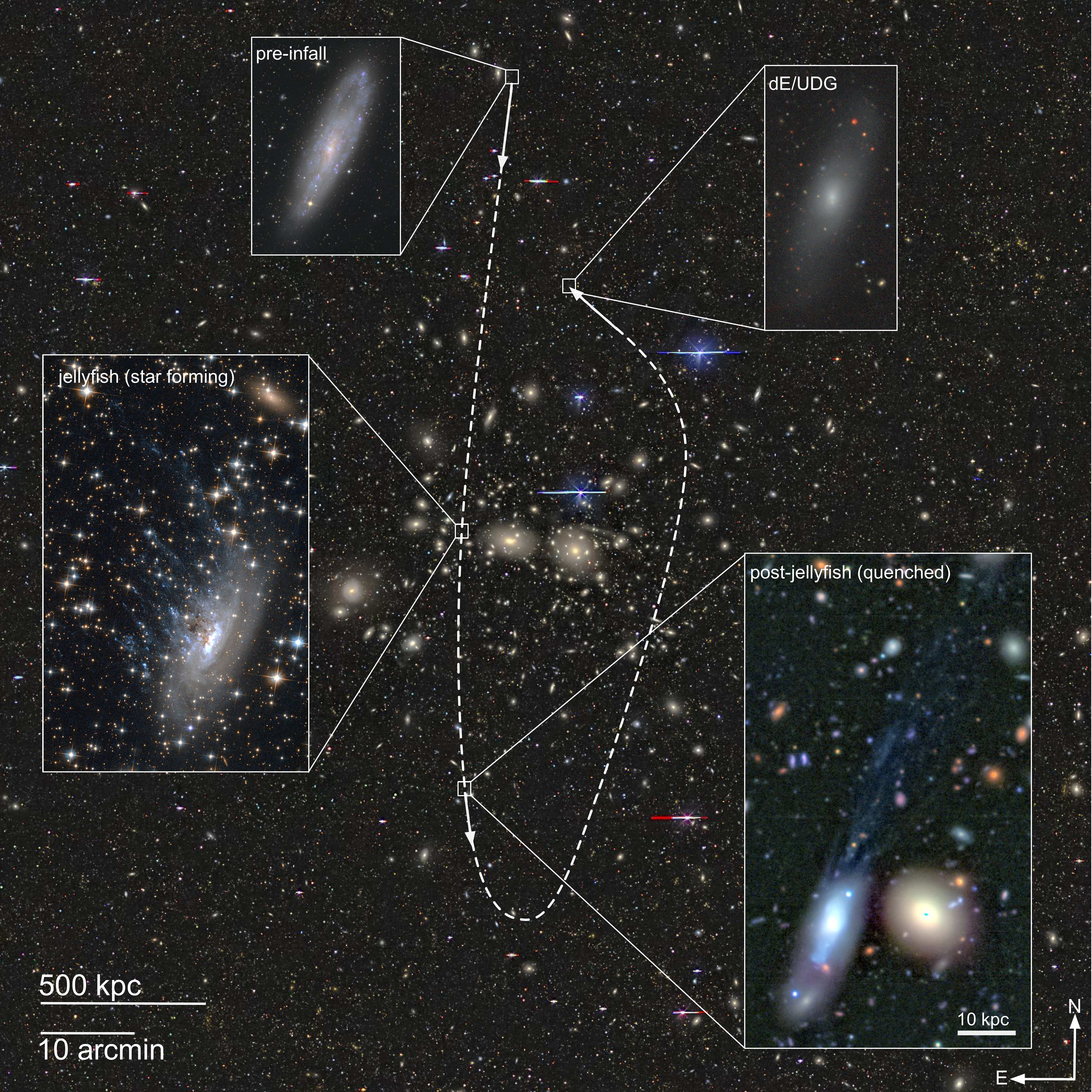}
    \caption{An artist's impression of an evolutionary path of a normal low-mass disk galaxy into a UDG/dE in a cluster. (a) An unperturbed galaxy falls into the cluster following an orbit with a substantial tangential component (shown by a dashed line). (b) Interaction with the intergalactic medium induces a burst of star formation and causes gas stripping, creating a `jellyfish' galaxy. (c) Star formation and stripping remove all of the gas, quenching further star formation and causing the disk expansion. (d) The galaxy evolves into a UDG or dE inside the cluster for billions of years.}
    \label{fig:udg_evo}
\end{figure}

Which phenomena are responsible for the trend of $M_*/M_{200}$ vs $M_{200}$?  Does this result from assumptions made during dynamical modelling? \citet{2021NatAs...5.1308G} demonstrated statistically that almost half the brightest UDGs in the Coma cluster were likely formed by ram pressure stripping of gas by the hot intracluster medium \citep{1972ApJ...176....1G} from low-mass disky progenitors. Moreover, prior to stripping, a burst of star formation is induced by compression of the gas inside an infalling galaxy, and up-to 50\%\ of the total stellar mass is formed during this event. Later on, gas from a galaxy gets stripped entirely, and star formation quenches. In a gas-rich galaxy, a substantial fraction of the mass can be stripped in the form of gas, so that the gravitational potential in the galactic disk changes, and quasi-circular orbits of stars become elliptical leading to the disk expansion. Later, the disk slowly expands even further as the galaxy passively evolves, losing stellar mass over time as the galaxy interacts with other members of the cluster. This process is illustrated in Fig.~\ref{fig:udg_evo}. 

The dark matter halo helps to stabilize the process of ram pressure stripping preventing a galaxy from a complete dissipation as a result of substantial mass loss.  The distributions of both dark matter and baryons can change as a result of mass loss, potentially affecting the dynamical modelling results from our spectroscopic observations ($<2\dots3 R_e$), where we have assumed a fixed halo density profile.

Another possibility is halo truncation from tidal interactions with other galaxies in the cluster. In this case the extrapolation of the dynamical properties derived from the `dense' baryonic part of the galaxy becomes invalid at $R_{200}$. However, we do not expect this process to be important for galaxies infalling into the cluster for the first time, such as the diffuse PSGs we observe.

\bibliographystyle{aasjournal_nodoi}
\bibliography{dE2023.bib}


\end{document}